\documentclass[referee]{aa}
\usepackage{graphicx}
\usepackage{txfonts}
\usepackage{cases}
\begin{document}
   \title{Observation and Simulation of Longitudinal Oscillations of an
	  Active Region Prominence}

   \author{Q. M. Zhang\inst{1} \and
           P. F. Chen\inst{1,2} \and
           C. Xia\inst{1} \and
           R. Keppens\inst{3}
          }

   \institute{School of Astronomy and Space Science, Nanjing University,
              Nanjing 210093, China
              \and
              Key Lab of Modern Astronomy and Astrophysics (Ministry of
              Education), Nanjing University, China\\
              \email{chenpf@nju.edu.cn}
              \and
              Center for Plasma Astrophysics, K.U.Leuven,
              Celestijnenlaan 200B, 3001 Heverlee, Belgium
              \\
              }

   \date{Received; accepted}
    \titlerunning{Evidence For Longitudinal Prominence Oscillations}
    \authorrunning{Zhang et al.}

  \abstract
  {Filament longitudinal oscillations have been observed on the solar disk
   in H$\alpha$.}
  {We intend to find an example of the longitudinal oscillations of a 
   prominence, where the magnetic dip can be seen directly, 
   and examine what is the restoring force of such kind of oscillations.}
  {We carry out a multiwavelength data analysis of the active region 
   prominence oscillations above the western limb on 2007 
   February 8. Besides, we perform a one-dimensional hydrodynamic 
   simulation of the longitudinal oscillations.}
  {The high-resolution observations by Hinode/SOT indicate that the 
   prominence, seen as a concave-inward shape in lower-resolution Extreme 
   Ultraviolet (EUV) images, actually consists of many concave-outward 
   threads, which is indicative of the existence of magnetic dips. After
   being injected into the dip region, a bulk of prominence material started
   to oscillate for more than 3.5 hours, with the period being 52 min. The
   oscillation decayed with time, with the decay timescale being 133 min.
   Our hydrodynamic simulation can well reproduce the oscillation period, but
   the damping timescale in the simulation is 1.5 times as long as the
   observations.}
  {The results clearly show the prominence longitudinal oscillations around
   the dip of the prominence and our study suggests that the restoring force
   of the longitudinal oscillations might be the gravity. Radiation and heat
   conduction are insufficient to explain the decay of the oscillations. Other
   mechanisms, such as wave leakage and mass accretion, have to be
   considered. The possible relation between the longitudinal oscillations
   and the later eruption of a prominence thread, as well as a coronal mass
   ejection (CME), is also discussed.}

   \keywords{Sun: filaments, prominences --
             Sun: oscillations --
             Methods: observational --
             Methods: numerical
             }

   \maketitle

\section{Introduction} \label{S-intro}

Solar prominences are cold ($\sim$10$^4$ K) and dense (10$^{10}$$-$10$^{11}$
cm$^{-3}$) plasma suspended in the hot corona. They appear to be dark
filaments in the H$\alpha$ images on the solar disk (Tandberg-Hanssen
\cite{tan95}). Prominences (or filaments) are formed above the magnetic
polarity inversion lines (Zirker 
\cite{zir89}; Martin \cite{mar98}; Berger et al. \cite{ber08}; Ning et al.
\cite{ning09}). It can be formed within several hours, experiencing from hot
plasma to cold condensation (Liu, Berger \& Low \cite{liu12}). It is
generally thought that the equilibrium of prominences is maintained by the
force balance between the gravity of prominence and the magnetic tension force
of the dip-shaped field lines, although prominences were seldom observed to
have a dipped shape. Theoretical models suggest that dips exist in two types
of magnetic configurations, one is of the normal-polarity (Kippenhahn \& 
Schl{\"u}ter \cite{kip57}), and the other is of the inverse-polarity (Kuperus
\& Raadu \cite{kup74}). The first type can be formed in sheared arcades with a
weak twist (Antiochos et al. \cite{ant94,ant99}; DeVore \& Antiochos 
\cite{dev00}; Aulanier \& Schmieder \cite{aul02}; Karpen \& Antiochos 
\cite{karp08}; Luna, Karpen \& Devore \cite{luna12}), and the second type has 
a helical flux rope with a stronger twist that may either emerge from the
subsurface (Lites \cite{lit05}) or be formed in the corona due to magnetic
reconnection (van Ballegooijen \& Martens \cite{vanb89}; Amari et al.
\cite{ama00}). 

Many prominences end up with a final eruption to become coronal mass ejections
(CMEs). The eruption is generally triggered by photospheric motions, emerging
magnetic flux, or internal reconnection (see Forbes et al. \cite{forb06} and
Chen \cite{chen11} for reviews). Whatever mechanism is involved in triggering
the eruption, the triggering process, as a kind of perturbation, would generate
waves and oscillations in the prominence. Therefore, based on this line of
thought and spectroscopic observations of an oscillating prominence prior to
eruption, Chen, Innes \& Solanki (\cite{chen08}) proposed that long-time
prominence oscillations can be considered as one of the precursors for CME
eruptions. Such a prominence oscillation prior to eruption was also observed
by Bocchialini et al. (\cite{bocc11}), and the oscillation can even sustain 
until the eruption phase (Isobe \& Tripathi \cite{iso06}; Gosain et al. 
\cite{gosa09}). The restoring force for this kind of transverse oscillations 
was generally thought to be the magnetic tension force.

In addition to the transverse oscillations which were widely investigated
(e.g., Lin et al. \cite{lin07}), prominences may also have longitudinal
oscillations. Jing et al. (\cite{jing03}) for the first time found that,
initiated by a subflare, a filament started to oscillate along its axis. They
mentioned several possibilities of the restoring force for the oscillations,
including the gravity and a reflecting Alfv\'en wave package. Vr{\v s}nak et
al. (\cite{vrn07}) studied a similar event, and they attributed the restoring
force to magnetic pressure gradient along the field lines. With
radiative hydrodynamic simulations, Luna \& Karpen (\cite{lu12}) claimed that
the projected gravity along the flux tube should act as the restoring force.
One issue with the past observations is that the longitudinal oscillations
were observed on the solar disk, where the shape of the flux tube cannot be
detected, so it is not possible to check whether the field-aligned gravity
component can explain the observed oscillation period. Therefore, it would be
of interest to investigate the longitudinal oscillations along a prominence,
where the shape of the flux tube can be inferred from the high-resolution
observations. It is also interesting to check whether longitudinal oscillations
may precede the eruption of the prominence.

In this paper, we report the longitudinal oscillations of an active region 
prominence, which was observed by Hinode (Kosugi et al. \cite{kos07})
satellite with a high resolution. The data processing is described 
in Section~\ref{S-data}, and the analysis is presented in 
Section~\ref{S-observe}. In Section \ref{S-simulation}, we perform a 
one-dimensional (1D) hydrodynamic numerical simulation to reproduce the damped
oscillations of the prominence. Discussions and a summary are presented  in 
Sections~\ref{S-dis} and \ref{S-summary}, respectively.

\section{Observations and Data Processing} \label{S-data}

\begin{figure}
\centering
\includegraphics[width=12cm]{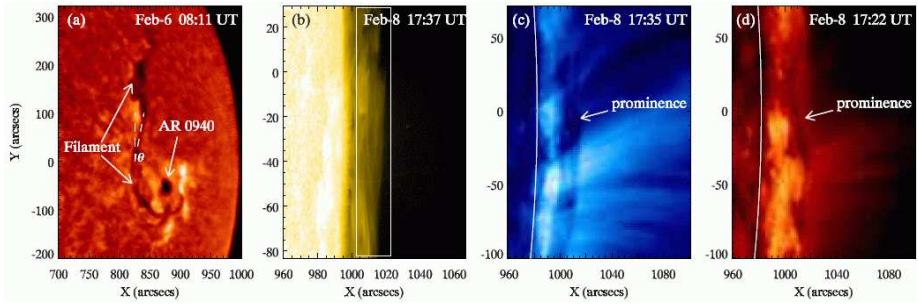}
\caption{Panel {\bf a)}: H$\alpha$ image at 08:11 UT on 2007 February 6
showing a filament near active region 0940; Panel {\bf b)}: Hinode/SOT 
\ion{Ca}{\sc ii} H image showing that the filament appeared above the
western limb as a prominence on February 8. The rectangle indicates a wide
slice used for Fig.~\ref{fig3}; Panel {\bf c)}: STEREO/EUVI 171 {\AA} image 
at 17:35 UT on February 8; Panel {\bf d)}: STEREO/EUVI 304 {\AA} image at 
17:22 UT on February 8. 
\label{fig1}}
\end{figure}

On 2007 February 6, there was a filament located in the NOAA Active Region
0940 (S05W71). The filament was seen to be slightly deviated from the
north-south direction in the H$\alpha$ images observed by Kanzelh\"ohe Solar
Observatory as shown in Fig.~\ref{fig1}a. It measured $\sim$300{\arcsec} in
length and 15\arcsec$-$20{\arcsec} in width. The southern part of the filament
resided in the active region, while the northern part extended far to the
quiescent region along the magnetic polarity inversion line. On February 8, as
the Sun rotated, the filament became a prominence and was monitored by the 
\ion{Ca}{\sc ii} H channel of the broadband filtergraph aboard the Hinode
satellite that carries three instruments: Solar Optical Telescope (SOT;
Tsuneta et al. \cite{tsu08}), X-Ray Telescope (XRT; Golub et al. 
\cite{gol07}), and EUV Imaging Spectrometer (EIS; Culhane et al. 
\cite{cul07}). Due to the limited field of view of SOT (111\farcs5 $\times$ 
111\farcs5), only a segment of the prominence was observed, which is shown in
Fig.~\ref{fig1}b. The pixel size and time cadence of the SOT observations are
0\farcs08 and 8 s, respectively. The observations started from 15:01 UT and
stopped at 21:24 UT. From $\sim$17:20 UT, a bulk of dense material was injected
from the south to the SOT field of view. The dense plasma began to oscillate
for 3.5 hours till the end of the observations.
The raw SOT data are calibrated using the standard Solar Software program 
\emph{fg\_prep.pro}. Note that a data gap exists from 18:00 UT to 18:14 UT. 

The prominence was also observed by Extreme Ultraviolet Imager (EUVI; 
Newmark et al. \cite{new07}) on board the Solar TErrestrial RElations 
Observatory (STEREO-A; Kaiser \cite{kai05}) spacecraft from a slightly 
different viewing angle. The 171 \AA\ and 304 \AA\ are shown in panels 
(c) and (d) of Fig.~\ref{fig1}, where we can
see that the prominence was straddled by much higher coronal loops that
are much hotter than the prominence. The pixel size and time cadence of 
the EUV observations are 1\farcs6 and 10 min, respectively. Its data 
calibration is conducted using the program \emph{secchi\_prep.pro}. 
Besides, the deviation of STEREO north-south direction from the solar 
rotation axis is corrected.

It is noted that there was a faint limb CME captured by the Large 
Angle and Spectrometric Coronagraph (LASCO; Brueckner et al. \cite{bru95}) 
aboard SOHO as well as the COR1 coronagraph aboard STEREO-A. It was registered 
by the CDAW CME catalog\footnote{http://cdaw.gsfc.nasa.gov/CME\_list/}
with a propagation speed of 480 km s$^{-1}$. The flare accompanying the CME
was registered as C1.2-class by the GOES satellite. The data from coronagraphs
were collected to investigate possible connection of the observed prominence 
dynamics to the CME.

\section{Observational Results}\label{S-observe}

\begin{figure}
\centering
\includegraphics[width=12cm]{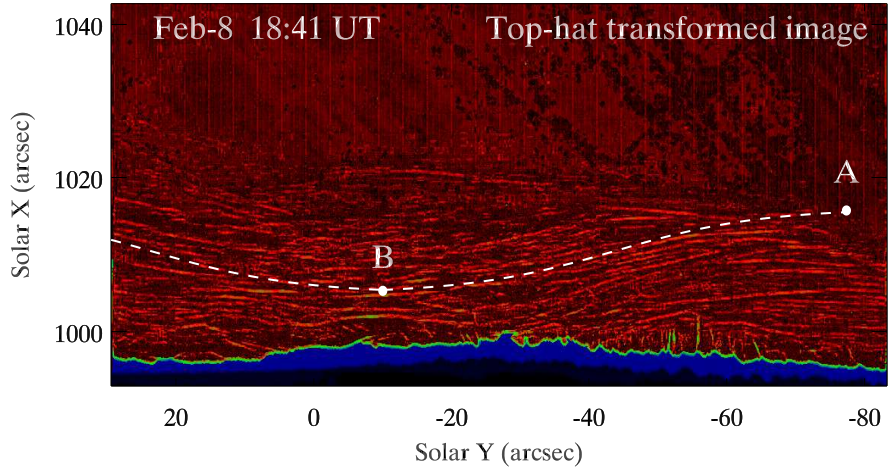}
\caption{Hinode/SOT Ca {\sc ii} H image at 18:41 UT on 2007 February 8, which
is processed with the top-hat transform. Points A and B mark the shoulder and 
the trough of a dipped prominence thread. Note that the image is rotated
counterclockwise by 90$^\circ$. North is to the left.
\label{fig2}}
\end{figure}

As displayed in Fig. \ref{fig1}, the prominence looked slightly different
in the chromospheric line \ion{Ca}{\sc ii} H and in the EUVI 304 {\AA} line.
While the lower-resolution 304 {\AA} image shows that the prominence was
a concave-inward structure at $-100\arcsec <y<0\arcsec$, the higher-resolution
\ion{Ca}{\sc ii} H image indicates that the prominence consists of a bunch of
concave-outward
threads. In order to show the dipped structure in \ion{Ca}{\sc ii} H more
clearly, we perform the top-hat transform of the Hinode image, which can
greatly enhance the detailed structures. Figure~\ref{fig2} displays the
processed Hinode \ion{Ca}{\sc ii} H image at 18:41 UT, where prominence
threads, as well as their oscillations, are found to be aligned with the
dipped trajectories. Since the \ion{Ca}{\sc ii} H thread is from the core
material of the prominence, which should trace the local magnetic field,
the concave-outward structure is strongly indicative of the existence of
magnetic dip along the prominence thread. The white dashed line traces one
single dipped thread, with point A located at one shoulder and point B at the
trough of an expected field line. The shape of this expected field line will
be used for numerical simulations in Sect. \ref{S-simulation}.

\begin{figure}
\centering
\includegraphics[width=12cm]{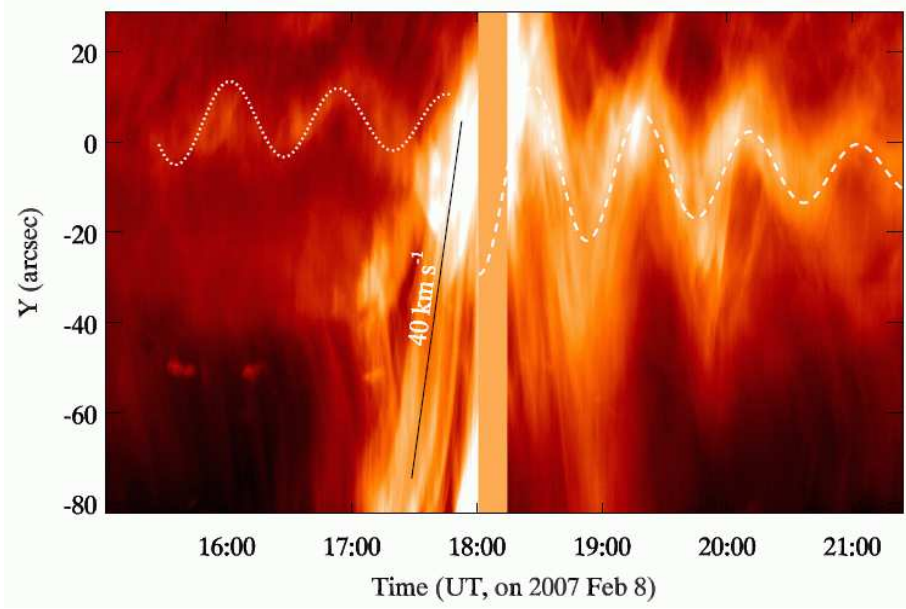}
\caption{Time-slice plot of the \ion{Ca}{\sc ii} H intensity along the
vertical direction of the rectangle in Fig.~\ref{fig1}(b). The intensity is
averaged along the horizontal direction. The superposed dashed line is a
damped sine function for fitting the mail oscillation, whereas the dotted line
for the weak prominence oscillation prior to the plasma injection. A data gap
exists between 18:00 UT and 18:14 UT.}
\label{fig3}
\end{figure}

At $\sim$17:00 UT, the prominence was activated somehow. In order to show the
dynamics of the prominence activation, in Fig. \ref{fig3} we plot the temporal
evolution of the \ion{Ca}{\sc ii} H intensity along a wide slice. The slice is
along the prominence axis in the south-north direction, with a width of 
20{\arcsec} as indicated in Fig. \ref{fig1}b. From Fig.~\ref{fig3} it can be
seen that at $\sim$17:00 UT a plasma clump formed near 
$y=-30\arcsec$, and then spread bi-directionally. During 17:10 UT--17:45 UT, a
bulk of dense plasma moved from south to north with a velocity of $\sim$40 km
s$^{-1}$. After a short interval, a denser plasma structure moved to
north. Although there was a data gap between 18:00 UT and 18:14 UT, it is still
discernible that whereas part of the dense plasmas moved outside the northern
edge of the Hinode/SOT field of view, the other part of them remained in the
field of view, and started to oscillate with an initial amplitude of 
$\sim$30\arcsec. The amplitude of
the oscillation decreased with time, but nearly four periods are visible in the
observation slot. In order to describe the prominence oscillation 
quantitatively, we fit the oscillating pattern in Fig.~\ref{fig3} with a
decayed sine function with respect to time, i.e.,

\begin{equation} \label{Eq-1}
y=A \sin(\frac{2\pi}{P} t+\phi) e^{-t/\tau}+y_0
\end{equation}
\noindent
where $A$ is the initial amplitude, $P$ the period, $\tau$ the decay
timescale, $\phi$ the phase, $t$ the time lapse since 18:00 UT, and
$y_0$ the equilibrium position of the prominence. By trial and error, we
found that with $A=24$ Mm, $\phi=-\pi/2$, $P=52$ min, and $\tau=133$ min, the
analytical function fits the oscillating pattern very well, as overplotted in
Fig.~\ref{fig3} with the dashed curve. The ratio of the decay timescale to
the period is 2.6, which is very similar to the results in Jing et al. 
(\cite{jing03}) and Vr{\v s}nak et al. (\cite{vrn07}).

It is noted that even before the bulk of dense plasma was injected into the
SOT field of view, as indicated by the dotted line in Fig. \ref{fig3}, the
less dense prominence material near $y=5^{''}$ was
already oscillating with almost the same period as the main oscillations
described above, which implies that the oscillation period was mainly
determined by the magnetic configuration, rather than the mass of the
prominence, as claimed by Luna \& Karpen (\cite{lu12}).

\section{Numerical Simulations} \label{S-simulation}

\subsection{Simulation setup} \label{S-method}

\begin{figure}
\centering
\includegraphics[width=12cm]{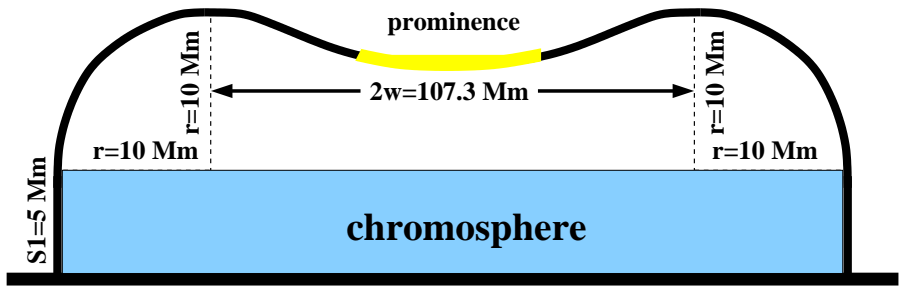}
\caption{Magnetic field configuration used for the 1D radiative hydrodynamic
simulation of the prominence oscillation. Note that the horizontal and the
vertical sizes are not in scale.}
\label{fig4}
\end{figure}

Jing et al. (\cite{jing03}) proposed several candidate restoring force of the 
oscillation, including the gravity and a reflecting Alfv\'en wave package, 
where the gravity acting as the restoring force was backed by Luna \&
Karpen (\cite{lu12}). On the other hand, Vr{\v s}nak et al. (\cite{vrn07})
attributed the restoring force to magnetic pressure gradient. Here, the fact
that the oscillation was centered around the magnetic dip reminds us of an
alternative explanation, i.e., the restoring force might be the gravity. In
order to explain the dynamics of a prominence after being disturbed, we 
performed a 1D radiative hydrodynamics simulation in this section. The
geometry of the dipped magnetic field line is taken from the observation in
Fig. \ref{fig2}, as marked by the white dashed line. Since the prominence
was inclined to the local meridian by $\theta=8^\circ$ as implied by Fig.
\ref{fig1}a when the prominence was observed on the solar disk as a filament,
the length of the dipped field line is $1/\cos(\theta+10^{\circ})$ times the 
apparent value assuming that the fine threads of the H$\alpha$ filament is
10$^{\circ}$ with respect to the filament axis. 
The reconstructed magnetic loop is depicted in Fig. \ref{fig4},
where the magnetic dip is 8.1 Mm in depth and 107.3 Mm in length.

The simulation setup was described in detail in Xia et al. (\cite{xia11, 
xia12}), and briefly explained as follows: A background heating is imposed
to sustain the corona-chromosphere structure. As an extra localized heating is
added at chromosphere symmetrically at two footpoints of the magnetic loop,
cool plasmas are heated up and evaporated into the corona. At a critical stage,
thermal instability is triggered in the corona, and the evaporated mass cools
down to form a condensation (or prominence segment). As the localized heating
is switched off, the condensation will relax to a quasi-static state. In order
to simulate the oscillation of the condensation, an impulsive momentum is
imposed on the plasma condensation, which makes the plasma to move with an
initial velocity of 40 km s$^{-1}$.

The 1D radiative and conductive hydrodynamic equations, as listed in Xia 
et al. (\cite{xia11}), are numerically solved with the Message Passing 
Interface$-$Adaptive Mesh Refinement Versatile Advection Code (MPI-AMRVAC, 
Toth \& Odstrcil \cite{toth96}; Keppens et al. \cite{kepp11}). The Total Variation
Diminishing Lax-Friedrichs (TVDLF) scheme using linear reconstruction and a
Woodward limiter are selected for the spatial differentiation, while the
predictor-corrector two-step explicit scheme is utilized for the time
progressing. Mesh refinement with six levels in a block-based approach is
applied.

\subsection{Simulation results}

\begin{figure}
\centering
\includegraphics[width=12cm]{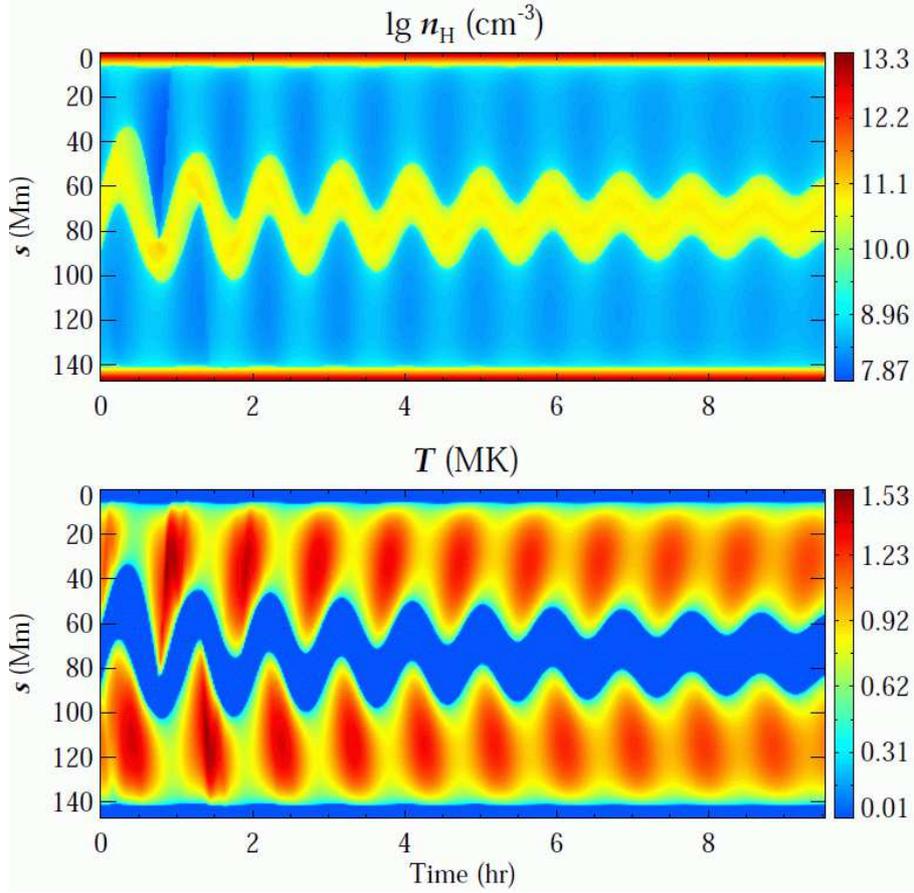}
\caption{Time evolutions of the density ({\it top}) and the temperature
({\it bottom}) distributions along the magnetic loop, which indicate that
the prominence experiences a damped oscillation subjected to a perturbation.
}
\label{fig5}
\end{figure}

As the initial conditions for the simulation in this paper, a prominence thread
formed due to thermal instability stays in a quasi-static state, with the
density being $\sim$2.0$\times 10^{11}$ cm$^{-3}$, the temperature being
$\sim$1.8$\times 10^{4}$ K, and the length being 27.8 Mm. A velocity
perturbation with an amplitude of 40 km s$^{-1}$ is introduced on the 
prominence thread, which decreases to zero smoothly near the 
prominence-corona transition layer. 

Subject to the velocity perturbation, the prominence thread begins to oscillate
along the dip. The time evolutions of the density and temperature of the plasma
along the magnetic loop are displayed in Fig.~\ref{fig5}, which shows that
the prominence thread oscillates with a damping amplitude. Using a formula
similar to Equation~(\ref{Eq-1}), we fit the displacement of the prominence
from the center of the dip as a function of time. It turns out that the initial 
amplitude is $A=18.6$ Mm, the period is $P=56$ min, and the damping timescale
is $\tau=202$ min. It is noticed that the oscillation period in the 1D
simulation, i.e., 56 min, is very close to the real observations. However, 
the decay timescale, 202 min, is $\sim$1.5 times as long as the real
observations between 18:00 UT and 21:24 UT.

Since the magnetic field line is fixed here in the 1D simulation, the
restoring force candidates in the simulation are the field-aligned component
of the gravity and the gas pressure gradient. After comparing the two forces
in the simulation results, we found that the gravity component is much larger
than the gas pressure gradient. Therefore, the result that the simulated
period is similar to the observed one indicates that the gravity component 
along the field line nicely accounts for the restoring force of the
longitudinal oscillation of the prominence. Our results are consistent
with the conclusion of Luna \& Karpen (\cite{lu12}) that the main restoring
force is the projected gravity in the flux tube dips where the threads
oscillate. The energy losing mechanisms
in our simulation include thermal radiation and heat conduction. The fact that
the decay timescale of the oscillation in our simulation is 1.5 times as long
as the observed one implies that some other energy loss mechanisms
should be taken into account, such as the wave leakage (e.g., Stenuit et al.
\cite{sten99}) and mass accretion (e.g., Luna \& Karpen \cite{lu12}).

\section{Discussions} \label{S-dis}

\subsection{Identification of the longitudinal oscillations}

Even a quiescent prominence is full of dynamics. The typical motion inside
prominences is the counter-streamings (Zirker et al. \cite{zir98}). Such 
bidirectional motions have been observed in quiescent region filaments
(Lin et al. \cite{lin05}; Schmieder et al. \cite{sch10}) as well as in 
active region prominences (Okamoto et al. \cite{oka07}). The counter-streaming
can be easily discerned in Fig. \ref{fig3} of this paper. Especially near
$t$=21:00 UT, counter-streamings are clearly seen around $y=-70^{''}$. With
such a background of counter-streamings, a distinct feature in Fig. \ref{fig3}
is that a bulk of prominence began to oscillate with a period of 52 min. The
oscillation decayed with a timescale of 133 min. It might be argued that the
oscillations would be an artificial pattern due to the superposition of the
on-going counter-streamings along the line of sight. We tend to discard such a
possibility and favor the oscillation explanation for several reasons: (1) The
counter-streamings are often random, and it is rare, if possible, for them to
form a coherent periodic oscillation pattern for $>$3.5 hours; (2) The
oscillation was centered
around the dip of the prominence, whereas counter-streamings appear everywhere
along the prominence axis; (3) The prominence oscillations in Fig. \ref{fig3} 
decayed in a steady way. It is a little hard to understand how the 
superposition of counter-streamings could evolve in such a systematic way;
(4) It is seen that even before the injection of the bulk of plasma, the less
dense prominence material near the dip was already oscillating with an
identical period to the main oscillations; (5) More importantly, with the 
shape of the magnetic flux tube inferred from observations, we performed
radiative hydrodynamic simulations, and the simulated oscillation period is so
close to the observed value, which strongly supports that the observed pattern
in Fig. \ref{fig3} presents the evidence of prominence longitudinal
oscillations, and the gravity serves as the restoring force for the
longitudinal oscillations.

It is also noted that only the portion of the prominence between $y=-40^{''}$
and $y=-20^{''}$ was oscillating, which is probably because only some threads
of the prominence were involved in the oscillation, and the other threads 
remained unaffected by the perturbation.

\subsection{Possible connection with the later CME/flare}

\begin{figure}
\centering
\includegraphics[width=12cm]{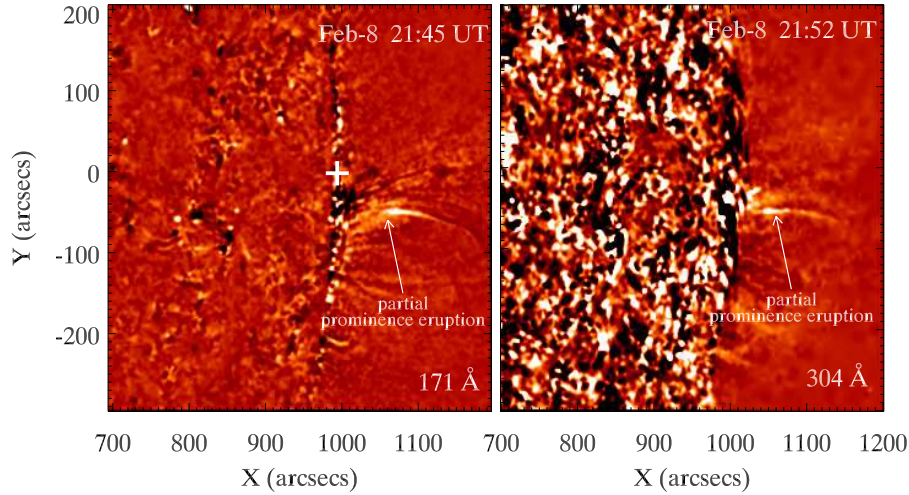}
\caption{The running-difference images of EUVI 171 {\AA} channel at
21:45 UT ({\it left}) and 304 {\AA} channel at 21:52 UT ({\it right}) showing
that only a thread of the prominence erupted. The white cross in the left
panel marks the location of a microflare which happened at 18:00 UT.
\label{fig6}}
\end{figure}

\begin{figure}
\centering
\includegraphics[width=12cm]{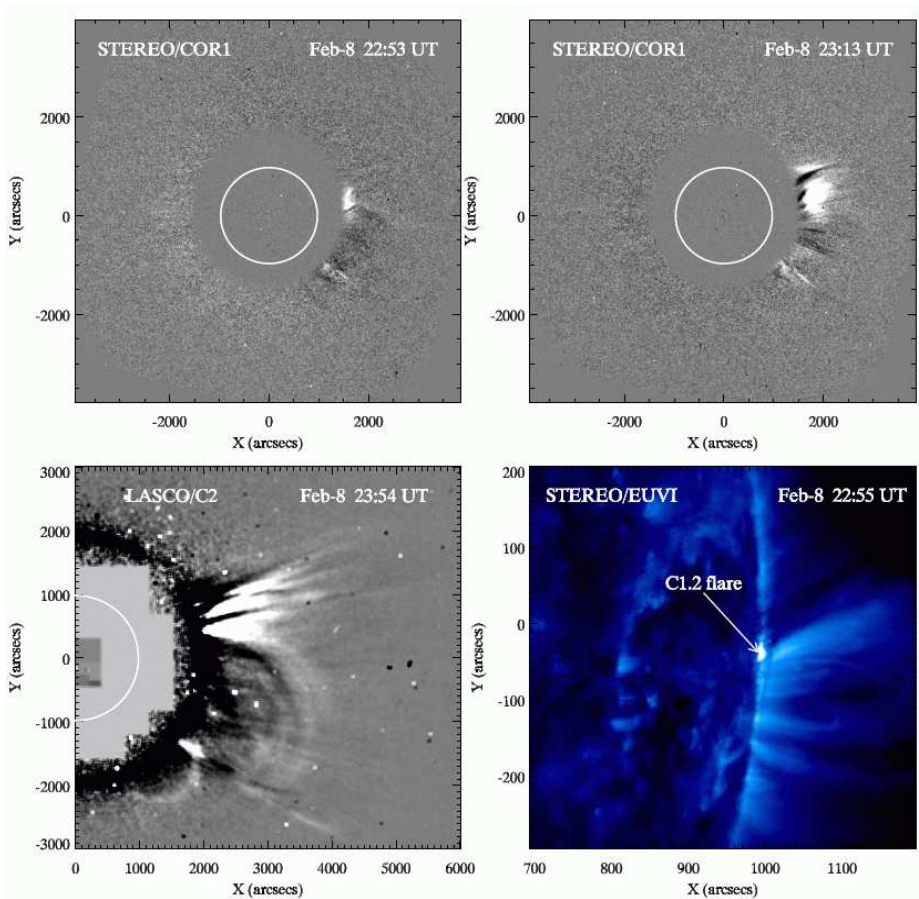}
\caption{\emph{Top}: STEREO/COR1 running-difference images showing
the early propagation of a CME; \emph{Bottom-left}: Running-difference
image from LASCO/C2 coronagraph at 23:54 UT; \emph{Bottom-right}:
\ion{Fe}{\sc ix/x} 171 {\AA} image at 22:55 UT showing a C1.2-class flare
associated with the CME.
\label{fig7}}
\end{figure}

No Hinode/SOT observations were available after 21:24 UT, while the 
prominence oscillation should still be going on. Then, around 21:52 UT, 
a thread of the prominence erupted, as illustrated by the EUVI 171 {\AA} 
and 304 {\AA} running-difference images in Fig. \ref{fig6}. In 
association with that, a faint CME first appeared in the field of view 
of the STEREO/COR1 coronagraph above the western limb at 22:53 UT, as 
illustrated by the top panels of Fig. \ref{fig7}. The faint CME was later 
observed by the LASCO coronagraph, which showed that the CME was 
deflected toward south as seen in the bottom-left panel of Fig. 
\ref{fig7}. In association of the CME, a C1.2-class flare appeared just
below the prominence, starting around 22:55 UT, as indicated by the
bottom-right panel of Fig. \ref{fig7}. It is noted that despite the 
large-scale CME eruption, the major part of the prominence remained at 
the original place. Although the available observations did not cover 
the evolutions of the prominence eruption and the CME continuously, we 
speculate the possible connection among the erupting prominence thread, 
the flare, and the CME with the following reasons: First, none of active 
regions or filaments existed west to AR0940 before February 9. Hence, the 
CME could not originate from any filament eruption on the backside of 
the Sun near the western limb. There was another long filament in AR0941 
east to AR0940. However, the He {\small I} 304 \AA\ movie reveals that
the filament did not erupt before the CME. Second, the central position 
angle of the CME is close to the latitude of the erupting thread of the 
prominence. Finally, as displayed below, the timing of all the components 
fits the standard CME/flare model so well, i.e., a prominence is somehow 
triggered to rise with the formation of a current sheet. The reconnection 
of the current sheet leads to a flare and the acceleration of the CME.

\begin{figure}
\centering
\includegraphics[width=12cm]{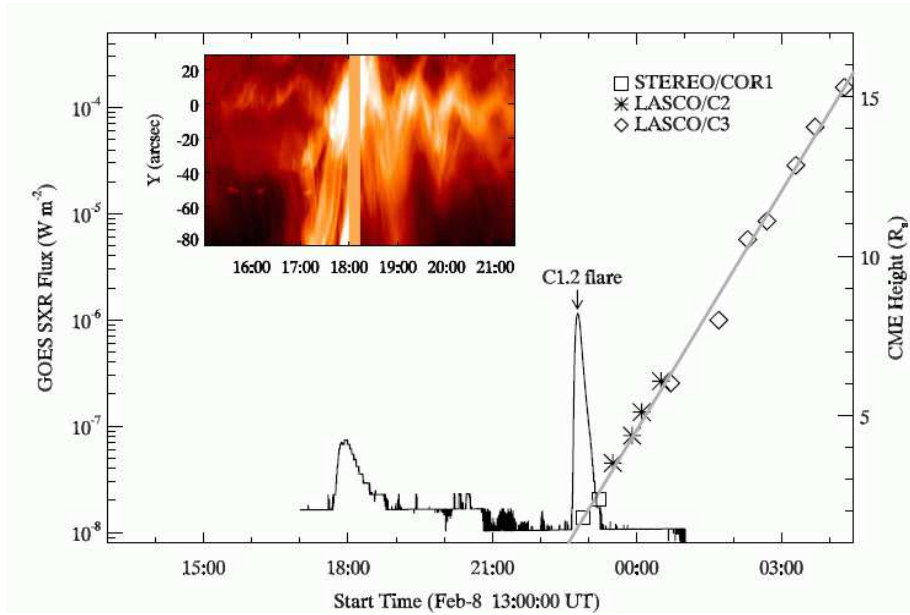}
\caption{Timeline of the whole event, where the
flare is indicated by the GOES 1$-$8 \AA\ SXR light curve ({\it solid
line}), the prominence oscillation is by the insertion of the time-slice plot
of Fig. \ref{fig3}, and the CME heights observed by different instruments,
i.e., STEREO/COR1 ({\it squares}), LASCO/C2 ({\it stars}), and LASCO/C3
({\it diamonds}), respectively.
\label{fig8}}
\end{figure}

The timeline of the whole event is illustrated in Fig.~\ref{fig8}. It is
seen that the prominence oscillation preceded the CME and the accompanying 
C1.2 flare. Despite that a data gap exists after 21:24 UT, we suspect that 
the longitudinal prominence could still keep oscillating for a few cycles more.
Since only one thread of the prominence was seen to erupt with the major part
remaining in the low corona, this should be a partial prominence eruption
event, which might be consistent with the rupture model (Sturrock et al.
\cite{stur01}), where part of the flux rope struggle out of the overlying
field lines, pushing some of them aside on the way, as simulated by Fan
(\cite{fan05}).

This paper might provide another example to illustrate that long-time
prominence oscillation can be regarded as a precursor for CME eruptions, as
proposed by Chen et al. (\cite{chen08}). However, the eruption in this paper
is different from theirs in the sense that their prominence oscillation was
transverse, whereas ours is along the prominence axis, i.e, it is a 
longitudinal oscillation, although both of the prominence oscillations were
probably triggered by magnetic reconnection between emerging flux and the
pre-existing field lines. Caveat should be taken that many of the CME
precursors are neither sufficient nor necessary conditions for the CME
eruption (Chen \cite{chen11}). Besides, in both Jing et al. (\cite{jing03})
and Vr{\v s}nak et al. (\cite{vrn07}), it seems that no CME followed the 
longitudinal oscillation of the prominence. 

Interestingly, an A7.0 microflare peaked
around 18:00 UT near the prominence, as seen from the GOES light curve in
Fig. \ref{fig8}. The location of the microflare is marked by the cross in the
left panel of Fig. \ref{fig6}. The microflare might be due to the reconnection
between emerging flux and the prominence magnetic field, which led to the
second prominence oscillation. However, caveats should be taken. First, this is a limb
event, so the microflare, which seemed to be close to the prominence in the
image might be far from the prominence. Second, the oscillation of the massive
prominence started $\sim$30 min before the microflare. It seems unlikely for
the microflare to trigger the prominence oscillation.

\section{Summary} \label{S-summary}

In summary, we carry out a multiwavelength data analysis and a radiative
hydrodynamic simulation of longitudinal oscillations of an active region
prominence along the axis on 2007 February 8. The main results are summarized
as follows:

1. Despite that the overall shape of the prominence is concave-inward, the
high-resolution observations by Hinode/SOT indicates that the prominence
consists of many threads which are actually concave-outward. The 
concave-outward structures are strongly indicative of the magnetic dips that
support the heavy prominence materials suggested by previous theoretical
models.

2. After being injected, a bulk of dense prominence plasma was seen to 
oscillate along the prominence main axis, i.e., a longitudinal oscillation
is discerned. The period and the damping timescale are 52 min and 133 min, 
respectively. The oscillations continued for more than 3.5 hours, and no
further observations were taken.

3. To testify the mechanisms of restoring force and amplitude decaying
of the prominence oscillations, we performed a 1D hydrodynamic numerical
simulation with the geometry of the dipped magnetic loop inferred from
observations. The oscillation period derived from the simulation is nearly
identical to the observed values, indicating that gravity might serve as the
restoring force for the prominence longitudinal oscillations, as mentioned
by Jing et al. (\cite{jing03}) and Luna \& Karpen (\cite{lu12}). However, the
decay timescale of the oscillation in the simulation, 202 min, is 1.5 times as
long as the observed value, 133 min, suggesting that mechanisms other than
thermal radiation and heat conduction, say the wave leakage and mass
accretion, might be more relevant for the energy loss during the prominence longitudinal oscillations.

4. With the main body of the prominence staying torpid, a thread from the
prominence was seen to erupt, leading to a CME and a C1.2-class flare.
We tentatively propose that the prominence longitudinal oscillations studied
in this paper might be a precursor of the CME/flare, which is related to the
triggering process of the eruption, as proposed by Chen et al. (\cite{chen08}).

\begin{acknowledgements}
The authors thank the anonymous referee for the valuable comments and
suggestions to improve the quality of this paper. 
Q. M. Zhang appreciates Z. J. Ning, M. D. Ding, C. Fang, and B. Vr{\v s}nak
for discussions and suggestions on this work. H$\alpha$ data are from
the Global High Resolution H$\alpha$ Network operated by the Big Bear Solar
Observatory, New Jersey Institute of Technology. Hinode is a Japanese Mission,
with NASA and STFC (UK) as international partners. STEREO/SECCHI data
are provided by a consortium of US, UK, Germany, Belgium, and France. The CME
catalog is generated and maintained at the CDAW Data Center by NASA and The
Catholic University of America in cooperation with the Naval Research
Laboratory. The research is supported by the Chinese foundations NSFC
(11025314, 10878002, and 10933003) and 2011CB811402. PFC thanks UCL/MSSL for
the hospitality during his stay.
\end{acknowledgements}

\end{document}